# High carrier mobility in transparent $Ba_{1-x}La_xSnO_3$ crystals with a wide band gap

X. Luo[1,2], Y. S. Oh[3], A. Sirenko[4], P. Gao[4], T. A. Tyson[4], K. Char[5],

and S-W. Cheong[1,2,3 a)]

[1] Laboratory for Pohang Emergent Materials, Pohang University of Science and Technology, Pohang 790-784, Korea

[2] Department of Physics, Pohang University of Science and Technology, Pohang 790-784, Korea

[3] Rutgers Center for Emergent Materials and Department of Physics & Astronomy, Rutgers University, Piscataway, New Jersey 08854, USA

[4] Department of Physics, New Jersey Institute of Technology, Newark, New Jersey 07102, USA

[5] Department of Physics & Astronomy, Seoul National University, Seoul 151-744, Korea

## Abstract

We discovered that perovskite $(Ba,La)SnO_3$ can have excellent carrier mobility even though its band gap is large. The Hall mobility of $Ba_{0.98}La_{0.02}SnO_3$ crystals with the n-type carrier concentration of $\sim 8\text{-}10\times 10^{19}$ cm$^{-3}$ is found to be $\sim 103$ cm$^2$ V$^{-1}$s$^{-1}$ at room temperature, and the precise measurement of the band gap $\Delta$ of a $BaSnO_3$ crystal shows $\Delta=4.05$ eV, which is significantly larger than those of other transparent conductive oxides. The high mobility with a wide band gap indicates that $(Ba,La)SnO_3$ is a promising candidate for transparent conductor applications and also epitaxial all-perovskite multilayer devices.

a) Author to whom correspondence should be addressed. Electronic mail: sangc@physics.rutgers.edu

Transparent conducting oxides (TCOs), exhibiting the contraindicative properties of optical transparency in the visible region and high DC electrical conductivity, are widely utilized as optical window electrodes in photovoltaic devices, liquid crystal displays and solar energy conversion devices.[1-3] There have been significant attempts to find alternative TCO materials to replace indium tin oxide due to its soaring price, and new TCOs such as doped ZnO and $SnO_2$ have been, in fact, already utilized. On the other hand, most of TCOs that have been investigated so far are associated with band gaps significantly less than 4 eV, so they do not transmit ultraviolet (UV) light. For example, the band gaps of $SrTiO_3$ (STO), ZnO, $In_2O_3$, and $SnO_2$ are 3.25, 3.3, 2.9, and 3.6 eV, respectively.[4-7] Thus, the discovery of TCOs with band gaps > 3.6 eV is central to enhance the efficiency of, for example, solar energy harvesting. Furthermore, the epitaxial all-perovskite multilayer heterostructures based on STO have recently attracted much attention due to the multiplicity of advantageous physical properties of perovskites. These all-perovskite multilayer heterostructures have great potentials for innovative micro- and nanoelectronic devices.[8, 9] However, the mobility of doped STO is low at room temperature (RT) (~11 $cm^2$ $V^{-1}s^{-1}$),[10] and the instability of oxygen content in doped oxides such as oxygen-deficient STO is often a detrimental issue inducing fatigue and degradation.[11] Therefore, the critical matter of the development of all-perovskite multilayer devices is finding new perovskite materials exhibiting high carrier mobility near RT and good oxygen stability.

Herein, we report that La-doped perovskite $BaSnO_3$ (BSO) exhibits high carrier mobility with a large band gap, and thus is a promising candidate for transparent conductor applications with possible use in epitaxial all-perovskite multilayer devices. Alkaline earth stannates, with the formula of $ASnO_3$ (A=Ca, Sr and Ba), are widely used in the electronic industry for their optimal dielectric and gas-sensing properties.[12-14] BSO has an ideal cubic perovskite structure, and is an insulator with a valence band derived from orbitals of π–symmetry (mainly of oxygen 2p-character) and a conduction band with dominant Sn 5s-character.[15] The reported value of band gap Δ varies over a wide range with a theoretical value of 0.79 eV,[15] and experimental values covering the range 3.23-4.02 eV.[16-18] Although bulk single crystals of (doped) BSO have never been described in literature, polycrystalline (poly) materials and epitaxial films of electron-doped BSO have been fabricated and characterized. Conduction with n-type carriers has been, indeed, realized in La or Sb-doped BSO, but the associated carrier mobility turns out to be minuscular: 0.69 cm$^2$ V$^{-1}$s$^{-1}$ or less.[18] We have grown bulk single crystals of (La-doped) BSO and have measured the comprehensive physical properties of (La-doped) BSO crystals including the precise value of the BSO band gap of 4.05 eV. The carrier mobility in Ba$_{0.98}$La$_{0.02}$SnO$_3$ crystals with the n-type carrier concentration of ~8-10×10$^{19}$ cm$^{-3}$ was determined (~103 cm$^2$ V$^{-1}$s$^{-1}$). This mobility value is comparable with that of doped ZnO bulk crystals (~205 cm$^2$ V$^{-1}$s$^{-1}$).[19] We have also found that the conductivity in La-doped BSO is highly stable against annealing in various atmospheres at high temperatures.

Poly samples of Ba$_{1-x}$La$_x$SnO$_3$ (BLSO) were synthesized using the appropriate amounts of BaCO$_3$, SnO$_2$ and La$_2$O$_3$ from $x$=0 to 0.04. Stoichiometric amounts of the reagents were mixed, calcined at 1500 °C for 20 hours in an alumina crucible, and then cooled down to RT. BLSO crystals were grown using a PbO-based flux. Crystals grew in the form of light reddish ($x$=0) or gray ($x$=0.01, 0.02 and 0.03) platelets with a typical size of ~3×2×1 mm$^3$. To find out the stability of oxygen content and conductivity, one poly La$_{0.02}$ was annealed at 1300 °C for 20 hours, one La$_{0.02}$ crystal was annealed at 1000 °C for 10 hours in N$_2$ gas flow, and another La$_{0.02}$ crystal was annealed at 700 °C for 5 hours in O$_2$ gas flow, followed by furnace cooling to RT. Powder X-ray diffraction (XRD) patterns were taken with Cu K$_{\alpha 1}$ ($\lambda$=1.5406) radiation using a PANalytical X′pert diffractometer. Electrical transport measurements were carried out using the four-probe method. Hall measurements were performed using the five-point method in a Quantum Design PPMS-9. Ellipsometry experiments were carried out at RT using a Woollams spectroscopic ellipsometer. For the sound velocity measurement, piezoelectric transducers (Pb(Zr,Ti)O$_3$ for longitudinal mode and X-cut LiNbO$_3$ for transverse mode) were bonded to two parallel faces of a BSO crystal. The temperature of the specimen was controlled using a PPMS-9. A generated pulse of a transducer with 10 MHz frequency traveled through the specimen, and was detected on the opposite face by the second transducer. The specific heat of a BSO crystal was also measured using a PPMS-9.

Figure 1(a) shows the x-ray diffraction (XRD) pattern of $La_{0.03}$ poly. The lattice parameters were obtained by fitting the XRD pattern using the FULLPROF program package.[20] The powder pattern can be indexed with the $Pm\bar{3}m$ space group. The cubic perovskite structure is shown in the Fig. 1(b). The fitted lattice parameter $a$ is 4.1181 (4) Å for $Ba_{0.97}La_{0.03}SnO_3$ poly, slightly larger than that of BSO. This increase, despite the fact that ionic radius of $La^{3+}$ is smaller than that of $Ba^{2+}$, originates likely from the decrease of Sn valence.[21] The variation of lattice parameter for $Ba_{1-x}La_xSnO_3$ (BLSO or $La_x$) is shown in Fig. 1(c). With substitution of La for Ba ions, the lattice parameter of BLSO shows a linear increase for $x$ changing from 0 to 0.03, in agreement with the standard Vegard's law. Diffraction peaks of an impurity phase ($La_2Sn_2O_7$) were found in the XRD pattern for $x$ = 0.04, so the lanthanum solubility limit in BLSO is about 3 % in our poly specimens.

Figure 2(a) shows the temperature dependence of the resistivity for BLSO poly specimens and crystals. The resistivity behavior of poly specimens is somewhat anomalous: [1] There exist only weak temperature dependences in the temperature range of 300 K and 2 K with magnitude comparable with those of low-doped semiconductors, [2] $La_{0.02}$ shows lower resistivity than those of $La_{0.01}$ and $La_{0.03}$, and [3] $La_{0.02}$ was annealed at 1300 $^o$C in $N_2$ gas flow to increase oxygen vacancies (thus, increase n-type doping level and possibly conduction). Annealing of $La_{0.02}$ did not induce any significant change of resistivity. Plausible conclusion from these observations is: BLSO does have a significant metallic-like conduction, but the conduction in poly specimens is dominated by grain boundary scattering. Furthermore, BLSO and its conduction are very stable in terms of oxygen stoichiometry. These possibilities are further supported by our experimental results on single crystals.

BLSO crystals exhibit much lower resistivity than poly specimens do even though all BLSO crystals and poly specimens show similar weak temperature dependence of resistivity. In addition, the resistivity of $La_{0.02}$ crystals is significantly lower than the reported optimum resistivity value of La-doped BSO thin films.[18] These results indicate that the grain boundary contribution dominates the measured resistivity in poly specimens. The possible presence of grain boundaries and/or dislocation boundaries in quasi-epitaxial BLSO films may be also responsible for the large resistivity in the films. We also found that $N_2$ annealing of a $La_{0.02}$ crystal at 1000 $^oC$ and $O_2$ annealing of a $La_{0.02}$ crystal at 700 $^oC$ change the overall resistivity little, revealing the good stability of oxygen stoichiometry and also conduction behavior. Note that there exist only small differences in conductivity among three $La_{0.02}$ crystals (as-grown, $N_2$ annealed and $O_2$ annealed), and there is no evident systematics in the differences, suggesting that the small conductivity differences in three crystals originate mostly from the compositional variation in different crystals. The resistivity values of BLSO poly specimens and crystals at RT vs. La composition $x$ are shown in Fig. 2(b). Our values of the resistivity for BLSO poly specimens match reasonably well with those of from Ref. 22. For the crystals, the resistivity values at RT are more than three orders of magnitude lower than those for the poly specimens. The lowest resistivity of $La_{0.02}$ crystal is $\sim5.9\times10^{-4}$ $\Omega$ cm at RT, and the corresponding conductivity $\sigma$ is $\sim1,690$ S cm$^{-1}$, which is higher than the industry standard for TCOs.[1]

Experiment on as-grown and $O_2$ annealed $La_{0.02}$ crystals at RT was performed to determine the Hall constant $R_H$ ($R_H=\partial\rho_H/\partial B$, where $\rho_H$ is the Hall resistivity and $B$ is magnetic field) and Hall mobility $\mu_H$. As shown in the inset of Fig. 3(a), the sign of $R_H$ is negative, indicating electron type of conduction carriers. The Hall mobility $\mu_H$ and the Hall carrier concentration $n_H$ of an $O_2$ annealed (as-grown) crystal are found to be $\sim$103 (78) cm$^2$ V$^{-1}$s$^{-1}$ and 10.2 (8.9)$\times10^{19}$ cm$^{-3}$, respectively. The mobility of our crystals is three orders of magnitude larger than the reported values on poly specimens ($\sim$0.1 cm$^2$ V$^{-1}$s$^{-1}$).[23] For comparison, our values of the Hall mobility and the Hall carrier concentration and the literature values for $Ba_{0.93}La_{0.07}SnO_3$ ($\sim$0.69 cm$^2$ V$^{-1}$s$^{-1}$ and 2$\times10^{21}$ cm$^{-3}$) and $Sr_{0.93}La_{0.07}SnO_3$ ($\sim$4.28 cm$^2$ V$^{-1}$s$^{-1}$ and 3.9$\times10^{20}$ cm$^{-3}$) thin films are shown in Fig. 3(a).[18, 24] Evidently, the Hall mobility of $La_{0.02}$ crystals is much larger than those of thin films. Note that the carrier mobility of $La_{0.02}$ crystals is comparable with those of doped ZnO bulk crystals (the highest reported mobility of ZnO at RT is ~205 cm$^2$ V$^{-1}$s$^{-1}$.[19]). On the other hand, the mobility of La$_{0.02}$ crystals at RT is significantly larger than those of doped BaTiO$_3$ (~2 cm$^2$ V$^{-1}$s$^{-1}$),[25] CaTiO$_3$ (~8 cm$^2$ V$^{-1}$s$^{-1}$),[26] and STO (~11 cm$^2$ V$^{-1}$s$^{-1}$) at RT,[10] which have been utilized as the primary building block for high-mobility oxide heterostructures.[8, 9] Thus, the observed high mobility in doped BSO crystals indicates that doped BSO has a great potential for the primary building block for all-perovskite high-mobility heterostructures.

The band gap of a BSO crystal was measured by ellipsometry experiments, which have been carried out at RT in the energy range between 0.75 eV and 5.9 eV. The reflection plane of the crystal was oriented parallel to the [100] crystallographic direction. The results of the measurements for the real and imaginary parts of the dielectric function, $\varepsilon_1$ and $\varepsilon_2$, are shown in Fig. 3(b). In the low-energy limit, $\varepsilon_1$ is close to 4.3, while $\varepsilon_2$ approaches zero. The dielectric function was modeled using three harmonic oscillators with the energies $E$ of 4.05 eV, 4.93 eV, 5.71 eV and a Penn gap at 7.8 eV. The parameters of the oscillators are summarized in Table I. The Penn gap, which is above the measured energy range, was included in the model to take into account all higher-energy optical transitions above 6 eV and to obtain a match to the low energy value of $\varepsilon_1$ (0.75 eV) =4.3. The latter is impossible if only three oscillators (4.05 eV, 4.93 eV and 5.71 eV) are included in the dielectric model. The lowest energy band gap at 4.05 eV has a significant broadening with $\gamma$=0.85 eV, which significantly exceeds $k_BT$. This broadening is most likely due to defects or impurity Pb from flux incorporated at the Ba site. The band gap 4.05 eV of our crystal is larger than any reported values.[15-18] Although the mobility of La-doped crystals is comparable with that of doped ZnO crystals, the band gap is significantly larger than that of ZnO bulk crystals (~3.3 eV) at RT.[5] Using multiple angles of incidence (AOI) between 65° and 85°, we found that both $\varepsilon_1$ and $\varepsilon_2$ are independent on the AOI within the errors of the measurements. This observation confirms that the electronic gaps are isotropic, as expected for the cubic symmetry of BSO crystals.

The mobility of doped semiconductors or insulators at high temperatures such as RT depends strongly on the characteristics of low-energy (acoustic) phonons since electron-phonon scattering dominates the carrier conduction at high temperatures.

Thus, we have measured the specific heat and sound velocities of BSO crystals to unveil the nature of acoustic phonons. Figure 4(a) shows the temperature dependence of longitudinal and transverse sound velocity of a BSO crystal. At RT, transverse and longitudinal sound velocities are 3,333 m s$^{-1}$ and 6,150 m s$^{-1}$, respectively.[27, 28] The sound velocity of a specimen is associated with the linear dispersion ($\omega = vq$) of an acoustic phonon mode. Using the Debye theory of lattice dynamics, the acoustic Debye temperature ($\Theta_D$), which represents the energy scale of phonons, is estimated by $\Theta_D = \frac{h}{k_B} \left[ \frac{3qNd}{4\pi M} \right]^{1/3} v$ ($N$ is the Avogadro number, $d$ the density, $M$ the molecular weight, $q$ the number of atoms in a molecule and $v$ the longitudinal or transverse sound velocity).[29] The $\Theta_D$ of transverse and longitudinal acoustic branches are 382 K and 705 K, respectively. The Debye temperature also characterizes the thermodynamic property of acoustic phonons. Based on the Debye theory for one longitudinal and two transverse acoustic phonon modes, the temperature dependence of specific heat is described as follows:[30]

$$C_v = 9qNk_B \left[ x \left( \frac{T}{T_{Dt}} \right)^3 \int_0^{T_{Dt}/T} \frac{x^4 e^x}{(e^x - 1)^2} dx + y \left( \frac{T}{T_{Dl}} \right)^3 \int_0^{T_{Dl}/T} \frac{x^4 e^x}{(e^x - 1)^2} dx \right]$$

where $T_{Dt}$ and $T_{Dl}$ are the transverse and longitudinal thermal Debye temperatures, respectively, and $x$ and $y$ are weights of phonon modes. Figure 4(b) represents experimental specific heat data of a BSO crystal and the best fitting result with $T_{Dt}$ = 330 K and $T_{Dl}$ = 1,365 K for $x$=2/3 and $y$=1/3, corresponding to two transverse and one longitudinal acoustic phonon modes. The phonon density of states (DOS) in ABO$_3$ perovskites is dominated by the heavy A site ions at low energies and also have significant contributions up to high energies from the O ions.[31] Hence while the low energy peak (D$_t$) in the phonon DOS can be well modeled by a simple Debye type model, the higher mode (D$_l$) are influenced by additional phonon modes in this qualitative fit. The thermal Debye temperature for the transverse mode is consistent with the acoustic Debye temperatures for the transverse mode obtained from sound velocity measurements. The $C_P/T$ vs. $T^2$ plot for a BSO crystal in the Fig. 4(b) inset shows the presence of very low linear electronic contribution to specific heat, consistent with a doped-semiconductor-type small number of carrier concentration. Note that the Debye temperatures of BSO are comparable with those of STO and ZnO,[10, 32] but significantly smaller than that of, e.g., graphene.[33, 34] The slight reddish color of our BSO crystals grown with PbO-based flux possibly indicates the presence of Pb impurities in crystals, so the intrinsic mobility of La-doped BSO can be even larger than the value (103 cm$^2$ V$^{-1}$s$^{-1}$ at RT) that we have measured. However, the modest value of the Debye temperature (especially the low energy of transverse phonon modes) will limit the ultimate value of the intrinsic mobility of La-doped BSO at RT.

In summary, we have succeeded in growing transparent conductive BLSO crystals. Ba$_{0.98}$La$_{0.02}$SnO$_3$ crystals show low resistivity of 5.9×10$^{-4}$ Ω cm and large carrier mobility of ~103 cm$^2$ V$^{-1}$s$^{-1}$ at RT. The band gap $\Delta$ of a BSO crystal is measured to be 4.05 eV, which is larger than those of any other TCOs. We also found that the conductivity in La-doped BSO is highly stable against annealing in various atmospheres at high temperatures. The high mobility with a wide band gap and the good oxygen stability indicates the great potential of BLSO for transparent conductor applications and all-perovskite heterostructure devices.

**Acknowledgements**

Work at Rutgers and NJIT was supported by DOE DE-FG02-07ER46382 and DOE DE-FG02-07ER46402. Ellipsometry measurements were carried out at the Center for Functional nanomaterials, Brookhaven National Laboratory, which is supported by DOE DE-AC02-98CH10886.

**Table I.** Parameters of the harmonic oscillators that have been used in the dielectric function model shown in Fig. 3(b).

|  | $E$ [eV] | $\gamma$ [eV] | Amplitude (units of $\varepsilon_1$) |
|---|---|---|---|
| Oscillator 1 | 4.05 | 0.85 | 0.14 |
| Oscillator 2 | 4.93 | 0.67 | 0.21 |
| Oscillator 3 | 5.71 | 0.65 | 0.08 |
| Penn Gap | 7.8 | 3.14 | 2.15 |

**Figure captions:**

**Fig.1** (Color online) (a) Observed (symbols) and calculated (line) powder XRD patterns for poly $Ba_{0.97}La_{0.03}SnO_3$ at RT. The green line shows the difference between the observed and calculated diffraction patterns. (b) The cubic perovskite crystal structure of BSO. (c) Cubic lattice parameter for poly BLSO measured at RT. Line is drawn as a guide to the eyes.

**Fig. 2** (Color online) (a) Temperature dependence of resistivity for poly specimens and crystals of BLSO. The bottom three curves show resistivity of as-grown (blue), $N_2$ annealed (orange), and $O_2$ annealed (black) $La_{0.02}$ crystals. The purple line represents the published $Ba_{0.93}La_{0.07}SnO_3$ thin film data from Ref. 18. (b) La concentration dependence of resistivity at RT for poly specimens and crystals of BLSO at RT. Data from Ref. 22 are also included. Lines are drawn as guides to the eyes.

**Fig. 3** (Color online) (a) Carrier concentration vs. Hall mobility of $Ba_{0.98}La_{0.02}SnO_3$ crystals (blue: as-grown, black: $O_2$ annealed) and the literature values for $Ba_{0.93}La_{0.07}SnO_3$ and $Sr_{0.93}La_{0.07}SnO_3$ thin films (from Refs. 18 and 24). The inset shows Hall resistivity (blue diamonds) at RT vs. magnetic field for an as-grown $Ba_{0.98}La_{0.02}SnO_3$ crystal. Black line is a fit to obtain the Hall constant. (b) The real and imaginary parts of the dielectric function $\varepsilon_1$ (red circles) and $\varepsilon_2$ (blue triangles) for a BSO crystal measured at RT from the (100) plane with AOI=75°. The results of the fit to the dielectric function model that consists of three oscillators at 4.05 eV, 4.93 eV, 5.71 eV and a Penn gap at 7.8 eV are shown with black solid curves.

**Fig. 4** (Color online) (a) Temperature dependence of longitudinal (blue) and transverse (red) sound velocity of BSO crystal. (b) Experimental specific heat (red open circle) and phonon specific heat of a BSO crystal estimated using the Debye model (black line). The inset shows the $C_p/T$ vs. $T^2$ plot for a BSO crystal. Red line is drawn as a guide to the eyes.

**Figures:**

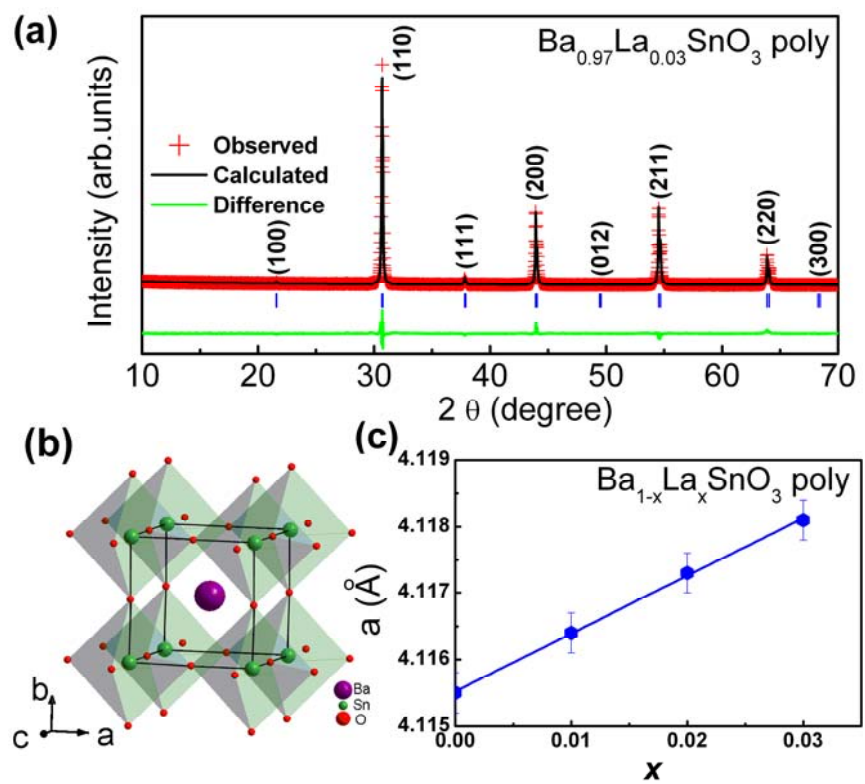

**Fig. 1 X. Luo *et. al.***

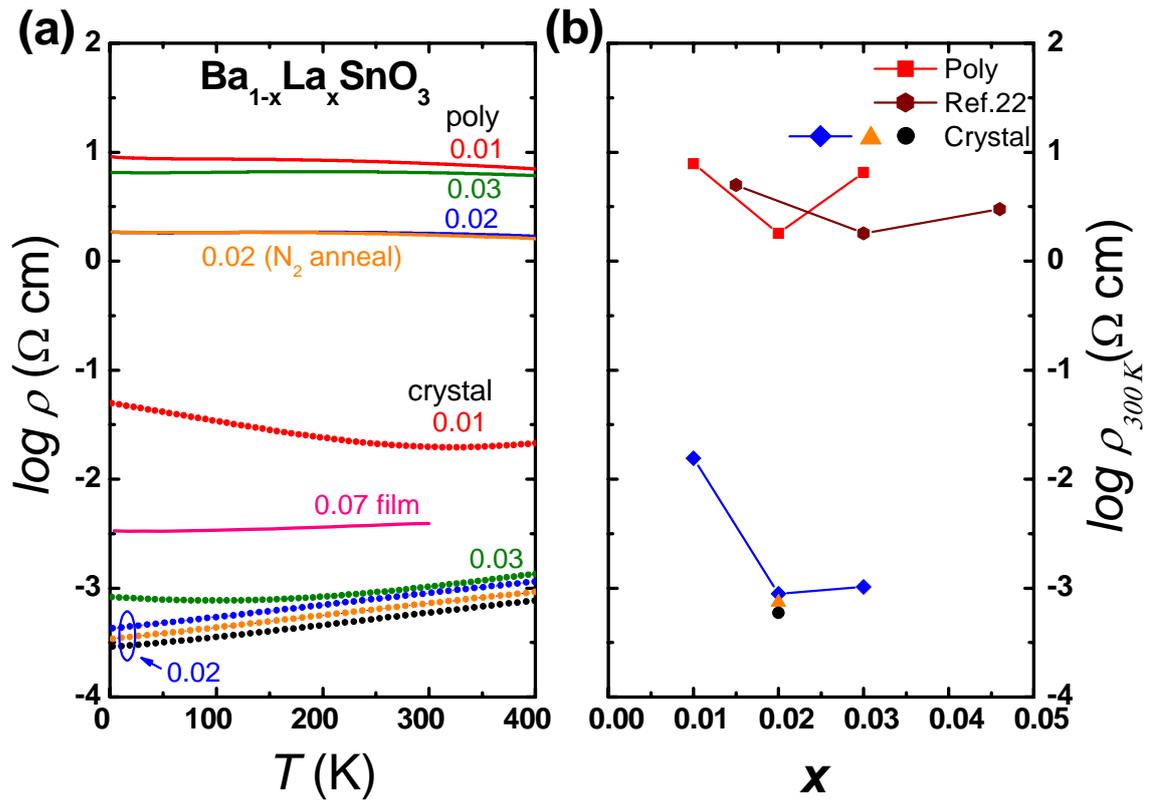

**Fig. 2** X. Luo *et. al.*

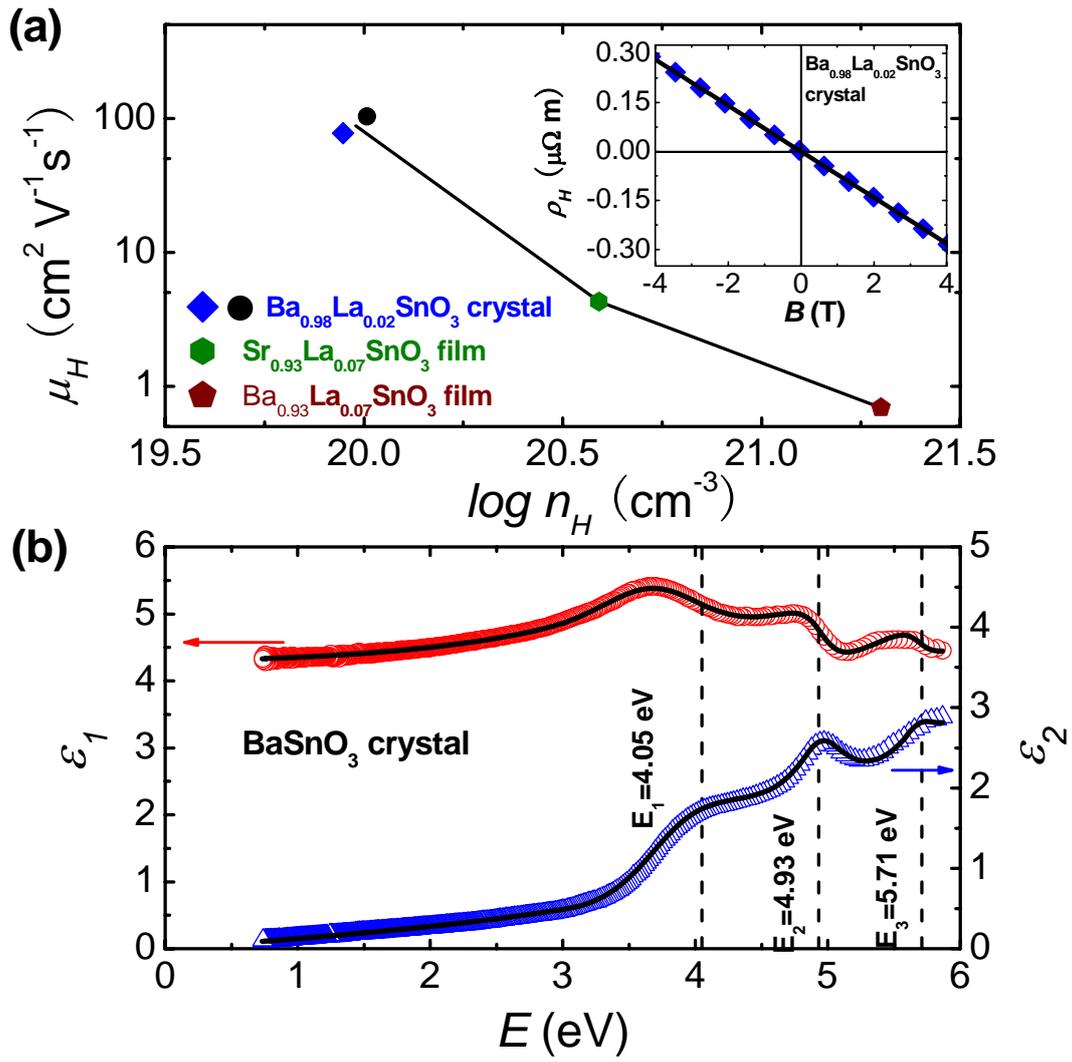

Fig. 3 X. Luo *et. al.*

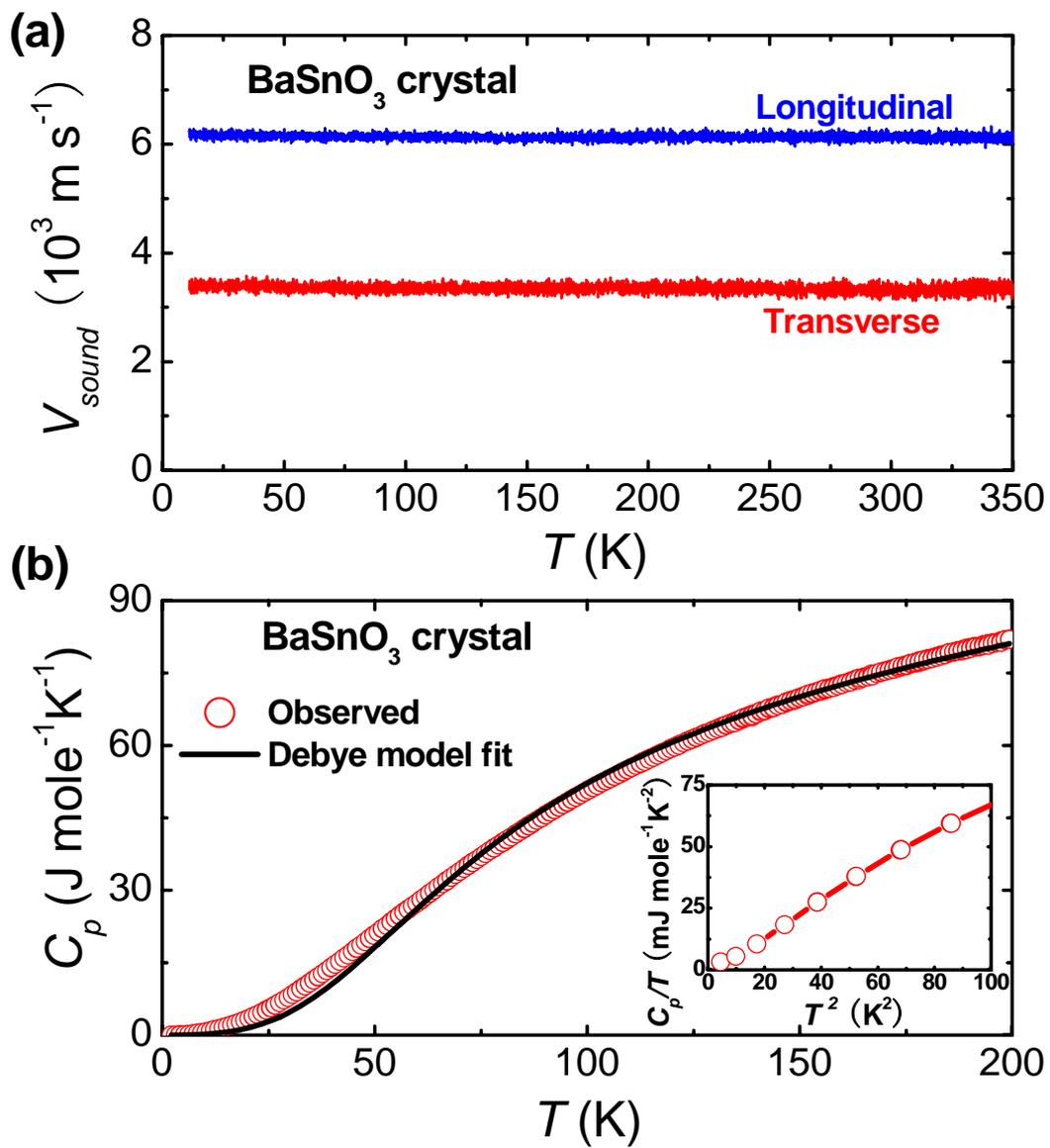

Fig. 4 X. Luo *et. al.*